\documentclass{webofc}
\usepackage[varg]{txfonts}   
\usepackage{multirow}
\usepackage{pstricks}
\usepackage{psfrag}
\usepackage{caption}
\newcommand{\MSb}{\overline{\mathrm{MS}}}

\begin{document}
\title{Overview of lattice calculations of the $x$-dependence of PDFs, GPDs and TMDs}

\author{\firstname{Krzysztof} \lastname{Cichy}\inst{1}\fnsep\thanks{\email{kcichy@amu.edu.pl}}
}

\institute{Faculty of Physics, Adam Mickiewicz University, \\
ul.\ Uniwersytetu Pozna\'{n}skiego 2, 61-614 Pozna\'{n}, Poland}
\abstract{%
  For a long time, lattice QCD was unable to address the $x$-dependence of partonic distributions, direct access to which is impossible in Euclidean spacetime. Recent years have brought a breakthrough for such calculations when it was realized that partonic light-cone correlations can be accessed through spatial correlations computable on the lattice. Appropriately devised observables can be factorized into physical PDFs via a perturbative procedure called matching, analogous to the standard factorization of experimental cross sections. In this short review, aimed at a broader high-energy and nuclear physics community, we  discuss the recent highlights of this research program. Key concepts are outlined, followed by a case study illustrating the typical stage of current lattice extractions and by a brief review of the most recent explorations. We finalize with a number of messages for the prospects of lattice determinations of partonic structure. 
}
\maketitle
\section{Introduction}
\label{intro}
Ever since the amazing discovery of nucleon's internal structure, it was one of the most important goals of hadron physics to fully understand and describe this structure.
For many years, the focus has been on longitudinal properties, especially for an unpolarized nucleon.
Several collider experiments can be well described taking only these properties into account.
They are summarized in unpolarized parton distribution functions (PDFs), objects that are well known from multiple global analyses based on thousands of experimental measurements.
However, it was realized that the full picture requires also addressing polarized nucleons -- thus, also polarized PDFs (helicity and transversity) are important.
These functions, in particular transversity, are significantly less constrained by collider data.
Moreover, description of some scattering events requires even more -- knowledge about the transverse position and motion of partonic constituents of hadrons.
Quantification of these aspects necessitates generalizing PDFs to more complicated distributions -- generalized parton densities (GPDs) and transverse-momentum-dependent PDFs (TMDs).
For these, our knowledge is even more restricted and we are basically at the beginning of the quest to unravel them.
This quest needs a coordinated effort of the experimental and theoretical/phenomenological communities.
The former will be enabled by the recent or ongoing construction of new experimental facilities, such as the 12 GeV upgrade at Jefferson Lab and the Electron-Ion Collider (EIC), with focus on providing the missing empirical data addressing fundamental questions about the nucleon structure.
In turn, the latter will consist in building realistic models, increasing the reliability of perturbative expansions, as well as attempting to access nucleon's properties from first principles.
This last possibility is the natural place where lattice QCD (LQCD) can be expected to yield progress, being the only feasible program to quantitatively extract QCD predictions directly from its Lagrangian.

Before calculations of $x$-dependent distributions became a reality, LQCD had a long history of computing moments of these functions, see e.g.~Ref.~\cite{Constantinou:2014tga} for a review of this period's achievements.
At that stage, around 2013-2014, the common belief was that LQCD was indeed restricted to calculations of moments, quantities accessible from local matrix elements (MEs) in Euclidean spacetime.
On the other hand, the full distributions, defined on the light front, were believed to be inaccessible without the Minkowski metric.
In principle, such distributions can be reconstructed from a tower of moments, but LQCD was understood to be practically limited to the lowest three, with higher ones suffering from inescapable power-divergent mixings with lower-dimensional operators and exponentially decaying signal-to-noise ratios.
Already at that stage, it was known that $x$-dependence could, in principle, be accessed even in Euclidean space, but the early proposals how to do it \cite{Liu:1993cv,Detmold:2005gg,Braun:2007wv} had not turned into practically feasible programs.
In 2013, Ji came up with another idea \cite{Ji:2013dva,Ji:2014gla}, known as quasi-distributions, to calculate appropriate spatial correlations in a boosted hadron instead of light-cone ones and relate the ensuing new objects with their physical counterparts via perturbation theory.
This is possible due to the asymptotic freedom of QCD and the quasi- and light-cone distributions having the same infrared (IR) properties.
As their differences emerge only in the ultraviolet (UV), one can use factorization and subtract the UV difference perturbatively, employing the so-called large momentum effective theory (LaMET).
This proposal immediately led to first lattice evaluations for the three twist-2 PDFs \cite{Lin:2014zya,Alexandrou:2015rja,Chen:2016utp,Alexandrou:2016jqi} and soon thereafter to a plethora of theoretical and practical studies.
The renewed interest in extracting the $x$-dependence on the lattice led to the revival of the earlier approaches and to the emergence of new ones \cite{Chambers:2017dov,Radyushkin:2017cyf,Radyushkin:2019mye,Ma:2014jla,Ma:2017pxb}, all of which have since achieved significant progress.
This dynamical progress is extensively reported in the reviews \cite{Cichy:2018mum,Ji:2020ect,Constantinou:2020pek,Cichy:2021lih}.

Since space for this proceeding is limited, it is not possible to discuss here all the progress even of the last year.
An interested Reader is, thus, referred to the above mentioned reviews, with the most recent developments covered in Ref.~\cite{Cichy:2021lih}.
Instead, we offer here a more general discussion addressed at a wider community, attempting to convey the key messages on what to expect from the lattice in the area of $x$-dependent distributions.

\section{Key concepts}
We start by reminding the Reader about the basics of the lattice formulation, to emphasize the implications for the robustness and reliability of lattice-extracted distributions.
LQCD is formulated on a discretized grid, with quarks living on lattice sites and gluons on links between them.
This grid is finite and discrete, thus IR (the lattice extent, $L$) and UV (the lattice spacing, $a$) regulators are naturally introduced.
Typical lattice sizes nowadays employed are $L/a\in[32,128]$ and lattice spacings $a\in[0.03,0.15]\,{\rm fm}$.
As quark masses are free parameters, one can simulate at arbitrary pion masses, typically chosen to be between the physical one and several hundred MeV.
The latter, while unphysical, allows one to significantly reduce the computational cost.
This cost is still rather huge -- the above setup regularizes the formally infinite QCD path integral, but the number of dimensions is still of $\mathcal{O}(10^8-10^9)$, necessitating the usage of world's most powerful supercomputers.

Obviously, the ultimate interest is in continuum QCD.
Consequently, the regulators need to be removed, as well as effects of other choices like the non-physical pion mass.
The key strength of LQCD is that all of this can be done in a controlled manner.
However, it also implies that lattice studies on a given subject proceed in stages:
\begin{enumerate}
\itemsep0pt
\item Feasibility studies.
\item Qualitative explorations.
\item Advanced studies.
\item Precision studies.
\end{enumerate}
The first stage only aims at establishing whether a given approach can be used in practice. Some methods may prove to be e.g.\ prohibitively expensive or may suffer from a very unfavorable signal-to-noise ratio. If the feasibility criterion is satisfied, the approach is typically explored qualitatively in a non-physical setup, e.g.\ at a large pion mass and with only a single lattice spacing.
Conclusions from such studies are usually only qualitative, with no quantitative comparison with experiment possible.
At this stage, possibities to optimize the method are also realized.
Later on, systematic effects are started to be addressed and first quantitatively reliable results can be obtained, although with rather large overall uncertainty.
In the last stage, additional optimizations of the method or setup can ensue and the fully quantified errors are reduced to a desired level, depending mostly on the available computer resources and uncertainties external to the lattice, such as truncation effects from a required perturbative input.

Needless to say, the duration of each stage is closely bound to the complexity of the physical question and the lattice method.
For the case of nucleon structure, in particular the $x$-dependent distributions, this complexity is significant and thus, the few years since Ji's breakthrough proposal and the revival/conception of other approaches have not been enough to get to the stage of precision studies.
For most approaches and distributions, the studies are somewhere between the first and third stage.
It is natural that studies of the simplest quantities, twist-2 PDFs, are most advanced -- quantification of systematics is progressing and in a few years, first precision studies may be expected.
The 3-dimensional distributions, in turn, lead to additional challenges and thus, are at a more exploratory stage.

Below, we discuss shortly the numerous aspects that need to be addressed to arrive at robust lattice-extracted distributions.
To make the arguments more concrete, we concentrate on the two most popular quasi- and pseudo-distribution approaches, in the context of calculating PDFs and GPDs.
In both of them, one computes MEs of the form 
${\mathcal M}_{\Gamma}(z,P)=\langle N(P_3) \vert \overline{\psi}(z)\Gamma \mathcal{A}(z,0)\psi(0)\vert N(P_3)\rangle$.
$\vert N(P_3)\rangle$ is a boosted hadron state, with boost $P_3$ taken in the $z$-direction.
A Wilson line of length $z$ in the same direction, $\mathcal{A}(z,0)$, is inserted to maintain gauge invariance.
The type of distribution is determined by the Dirac structure $\Gamma$.
In general, one can consider also momentum transfer, $t$, with $t=0$ ($t\neq0$) for PDFs (GPDs).
At $t\neq0$, it is also necessary to use additional projectors to disentangle the two (four) GPDs in the chiral-even (chiral-odd) case.
An important aspect is also that the spacetime locations of the source, the sink and the operator insertion need to be sufficiently separated to isolate the desired ground-state hadron.

The calculation of the above bare MEs is the most expensive part of the calculation.
It is important to emphasize that the actual cost may vary even by orders of magnitude, depending on simulation parameters, in particular the nucleon boost and the pion mass.
Increasing the nucleon boost leads to an exponentially decaying signal and the problem becomes even more severe when lowering the pion mass.
Working with large nucleon boosts (e.g.\ of $\mathcal{O}(3\,{\rm GeV})$) at the physical point, the signal is clean only at very low separations between the source and the sink.
However, in such a situation, one can hardly extract any ground-state properties, as the correlators receive sizable contributions additionally from tens of excited states, see Ref.~\cite{Green:2018vxw} for an extensive discussion.
The signal is further contaminated with noise when extracting GPDs, due to the nonzero momentum transfer and the necessity to use additional projectors.
An important technique to deal with the worsening noise for boosted hadrons is momentum smearing \cite{Bali:2016lva}.
It does not cure the exponential problem, but moves it to larger momenta.
Realistically, the signal-to-noise problem becomes severe when getting to boosts of order 2-2.5 GeV at typically used non-physical pion masses and already around 1.5 GeV at the physical point, see Tab.~1 of Ref.~\cite{Constantinou:2020pek} for a comparison of experience of different collaborations.
Larger boosts most likely suggest lack of control over excited states and unknown, possibly very large, systematic errors.
It should be kept in mind that large boosts also lead to enhanced discretization effects when the boost in lattice units becomes comparable to the inverse lattice spacing, the UV cutoff of the lattice theory.

Apart from the above, the calculation of bare MEs is subject to several standard sources of systematic effects.
In early stages of a computational program, these may be largely unexplored.
Addressing these systematics is mandatory towards the more mature stages and good control over them is a prerequisite for precision physics.
Thus, a fully controlled calculation necessitates the use of at least 3 lattice spacings to perform the continuum limit extrapolation (after removal of divergences), a check of finite volume effects, pion mass effects (if not simulating directly at the physical point) and several other sources of possible contaminations.
Depending on the final desired accuracy, one may need to address also subleading effects, e.g.\ the assumption that up and down quarks are degenerate is totally safe when aiming at 10\% precision, but cannot be made if 1\% total error is targeted; in such a case, one would need to take into account the different masses and electric charges of the light quarks.

The difficulty of the partonic distributions program consists also in the presence of additional sources of systematic effects beyond the ones for bare MEs.
An important issue is the non-perturbative renormalization of the latter.
The extra complication with respect to standard lattice hadron structure calculations arises due to the non-locality of inserted operators.
In approaches based on the Wilson line, its presence induces a power divergence, in addition to the standard logarithmic one.
The most commonly used renormalization strategies are types of RI/MOM renormalization tailored for non-local operators \cite{Constantinou:2017sej,Alexandrou:2017huk,Stewart:2017tvs,Izubuchi:2018srq} and ratio schemes \cite{Orginos:2017kos,Izubuchi:2018srq}.
However, as argued in Ref.~\cite{Ji:2020brr}, contamination from IR effects may ensue at large $z$.
In particular, RI/MOM schemes were shown to evince a residual divergence \cite{Zhang:2020rsx}, important numerically at very fine lattice spacings.
Proposals to overcome this issue were put forward recently, the so-called hybrid scheme \cite{Ji:2020brr} and ``self-renormalization'' \cite{LatticePartonCollaborationLPC:2021xdx}.

Further important systematics go beyond bare and renormalized MEs.
Since these MEs are Euclidean quantities, it is unsurprising that a delicate step is to ``translate'' Euclidean MEs to physical $x$-distributions.
This has two separate aspects, related to the change from coordinate to momentum space and from spatial to light-cone correlations.
At this stage, also the crucial difference between quasi- and pseudo-distribution approaches emerges.
In the former, renormalized MEs are first subjected to $x$-dependence reconstruction, leading to momentum-space quasi-distributions that are matched to their light-cone counterparts.
In turn, matching to physical distributions for the latter is performed in coordinate space, defining light-cone Ioffe-time distributions, then subjected to $x$-dependence reconstruction.
The reconstruction issue, broadly discussed in Ref.~\cite{Karpie:2019eiq}, originates from the mathematically ill-defined attempt to obtain a continuous distribution from a discrete and truncated set of MEs, which poses a so-called inverse problem.
The problem can be tackled by providing an additional input that complements the lattice data, in the form of a physically-justified assumption or a mathematical criterion.
As such, the solution of the inverse problem is non-unique and different methods should be used to avoid introducing a bias.
The translation from spatial to light-cone correlations (matching) is also subtle.
It can be done safely only if MEs were extracted for a sufficiently highly boosted nucleon.
In the presence of the exponentially decaying signal, care is needed to estimate effects from a finite boost.
These higher-twist effects are power-suppressed, but may be sizable with currently attainable hadron momenta and are expected to be enhanced both at small and large $x$.
Since matching relies on factorization, the involved perturbative expansion is also potentially sensitive to truncation effects.
For a long time only 1-loop formulae were known, but recently first calculations at two loops appeared \cite{Braun:2020ymy,Chen:2020arf,Chen:2020iqi,Chen:2020ody,Li:2020xml}, leading to conclusions about the size of truncation effects.
The latter need to be controlled also at the (separate or concurrent with matching) stage of translating from a lattice-applicable scheme to $\MSb$.

The above discussion has still not enumerated all possible sources of systematics. More elaborate considerations can be found in the reviews \cite{Cichy:2018mum,Ji:2020ect,Constantinou:2020pek,Cichy:2021lih} and in papers referenced therein.
Let us summarize this part with the main message of this discussion -- the lattice extraction of partonic distributions is comparatively more complicated with respect to most other lattice calculations and thus, subject to several other sources of systematic effects.
At the present stage, there is no complete calculation for any distribution in the sense of robustly quantifying all systematics.
As such, the current results are qualitative and no quantitative comparison with available experimental/phenomenological results should yet be attempted.
However, it is important to bear in mind that the sources of systematics are known and thus, they can and will be addressed in the rather near future.
In the next section, the above will be illustrated more concretely with recent lattice computations.

\section{Case study: twist-2 PDFs}
\label{sec:pdfs}
\begin{figure}[h!]
\begin{minipage}{0.35\textwidth}
\includegraphics[width=0.88\linewidth,keepaspectratio]{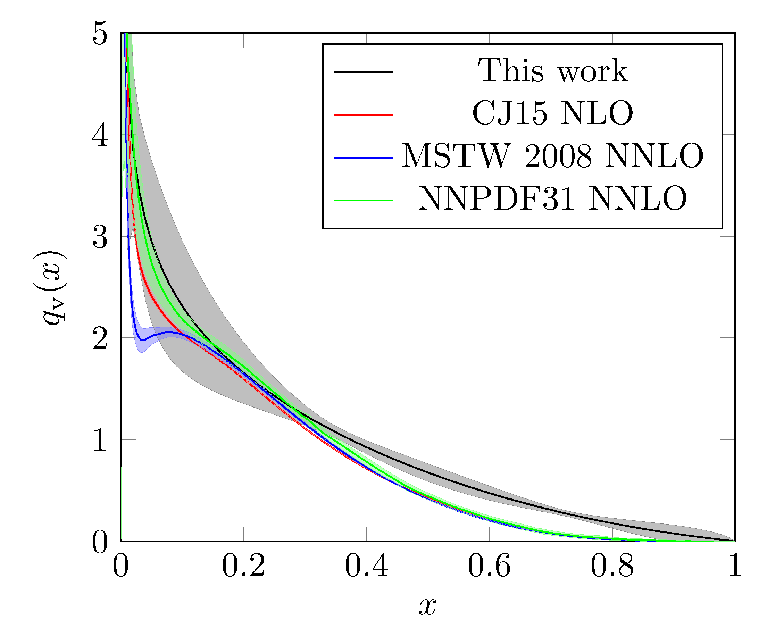}
\caption{\footnotesize\label{fig:unpol} Recent results for the nucleon's unpolarized isovector PDFs. Left: $q_v$ \cite{Joo:2020spy}. Right: $q_v$ (upper left), $q_{v2s}$ (upper right), $q$ (lower left), $\bar{q}$ (lower right) \cite{Bhat:2020ktg}. Plots reprinted based on the arXiv distribution license.}
\end{minipage}\hfill
\begin{minipage}{0.635\textwidth}
\hspace*{1mm}
\includegraphics[scale=0.425, angle=0]{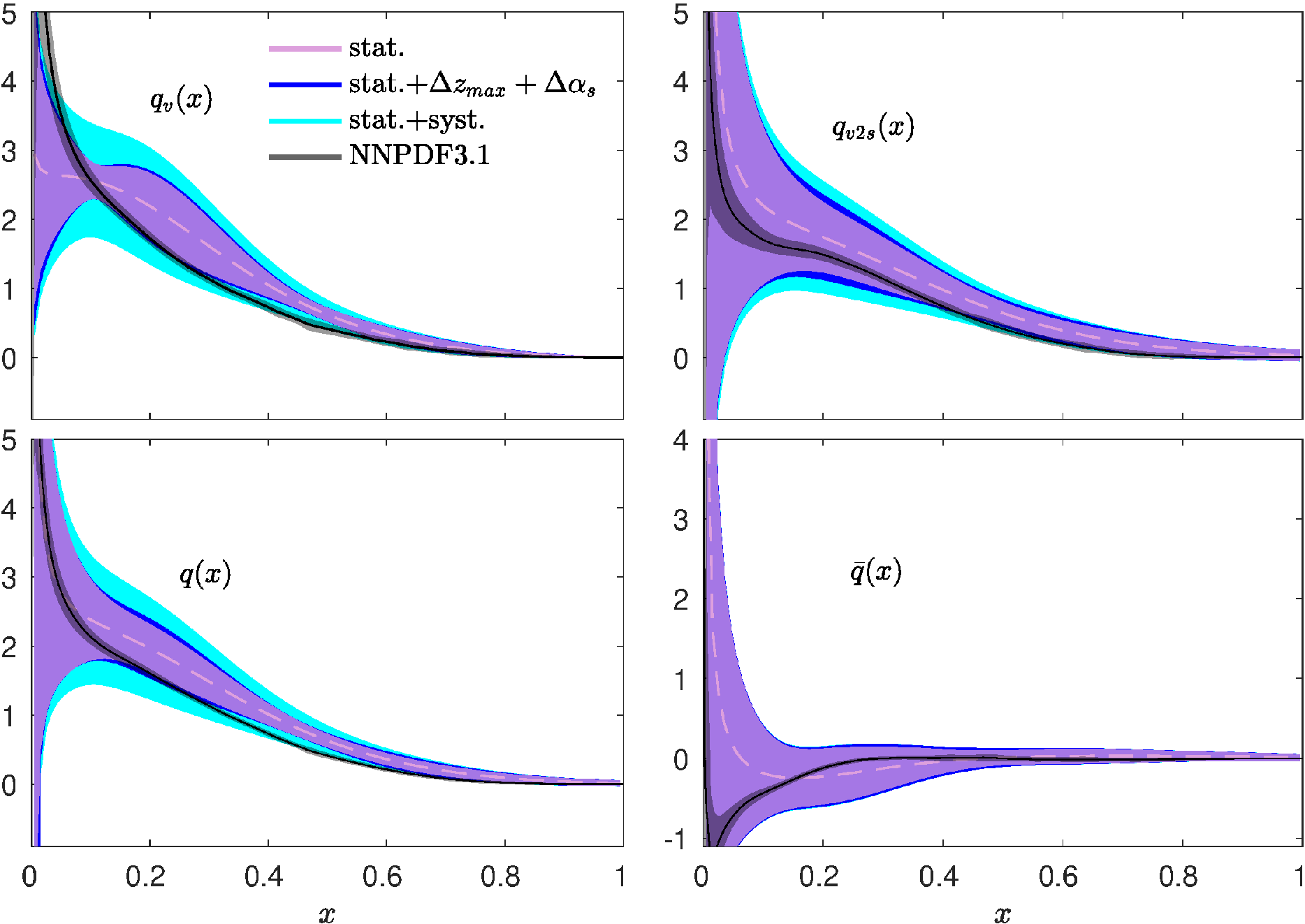}
\end{minipage}
\end{figure}

Unpolarized twist-2 isovector ($u-d$) PDFs are the simplest nucleon PDFs that can be extracted on the lattice and at the same time, the most well-known from phenomenological global fits.
As such, they are the natural benchmark case for LQCD and their accurate reproduction is an important aim to demonstrate the robustness of the lattice approach and good control over the numerous systematic effects.
The current state-of-the-art can be well illustrated based on two recent calculations at or near the physical point, from Extended Twisted Mass (ETMC) and HadStruc collaborations, using the pseudo-distribution approach.
The ETMC work \cite{Bhat:2020ktg} reused the data of Refs.~\cite{Alexandrou:2018pbm,Alexandrou:2019lfo}, where unpolarized and helicity PDFs were extracted in the quasi-distribution formalism from a physical point calculation using twisted mass (TM) fermions with a clover term, at a lattice spacing $a=0.094\,{\rm fm}$ and $L/a=48$.
In turn, HadStruc \cite{Joo:2020spy} used clover fermions at three pion masses, $m_\pi\!=\!358,278,172\,{\rm MeV}$, with the same lattice spacing and $L/a=32,64$.
The final distributions are shown in Fig.~\ref{fig:unpol}, the valence PDF $q_v$ for HadStruc and additionally different combinations with antiquarks for ETMC, $q_{v2s}\equiv q_v+2\bar{q}$, $q\equiv q_v+\bar{q}$ and $\bar{q}$ alone.

Direct comparison is possible for $q_v$.
Results from both calculations are qualitatively very similar. 
In HadStruc's work (left part of Fig.~\ref{fig:unpol}), there is good quantitative agreement with global fits for $x\lesssim0.4$ (where the errors are the largest) and a discrepancy with phenomenology observed at larger $x$.
This is the typical situation of present lattice extractions and it has a clear origin in unaccounted for systematics from various sources -- the plotted errors are merely statistical and do not reflect that the calculation was done only with a single lattice spacing, using only 1-loop matching etc.
Conclusions are similar also for ETMC (right part of Fig.~\ref{fig:unpol}) when only statistical errors are considered (purple bands).
The tension with global fits is somewhat smaller, which suggests an accidentally more favorable interplay of unquantified systematics.
However, in this case, an estimate of these systematics was attempted, taking typical (``plausible'') magnitudes (based on other hadron structure studies) of cutoff, finite volume, excited states and truncation effects, as proposed in Ref.~\cite{Cichy:2019ebf}, with two additional systematic effects assessed more quantitatively.
The enlarged total uncertainty from combining statistical and ``plausible'' systematic effects (cyan band) led to total agreement with global fits for the whole $x$-range.
While such a procedure for systematics cannot be treated as replacement for their careful (and laborious) quantification, it is an important check that systematic effects typically found in other studies may be enough to explain the remaining differences between lattice-extracted PDFs and their phenomenological counterparts.

The same lattice data were also subjected to analyses in the robust NNPDF (Ref.~\cite{Cichy:2019ebf} for ETMC and Ref.~\cite{DelDebbio:2020rgv} for HadStruc) and JAM (only for ETMC \cite{Bringewatt:2020ixn}) global fitting frameworks.
We only briefly summarize the main findings.
The NNPDF study of ETMC data (in the quasi-PDF formalism) led to conclusions similar to the above, with larger differences with respect to global fits.
This is unsurprising, since systematic effects are different at some stages of pseudo- and quasi-PDF extractions, here with their less favorable interplay in the latter.
However, again, the ``plausible'' magnitudes for unquantified systematics brought the final neural-network-reconstructed distributions to agreement with NNPDF3.1.
An estimate of systematics was performed also for the HadStruc data combined with the NNPDF framework, leading to better agreement with phenomenology as compared with the case of only statistical errors, but still with some leftover tensions due to incompleteness of systematic estimates at this stage.
Finally, the JAM framework as applied to ETMC unpolarized data pointed to similar conclusions, again confirming the need for careful systematic analyses and the need to reduce the overall uncertainties.
Interestingly, the JAM work also used lattice data for helicity PDFs.
This case is different from the phenomenological point of view, as this kind of PDFs is significantly less constrained by experimental data.
The authors concluded that the lattice data can already have some additional constraining power and the tensions with global fits are already smaller than for unpolarized PDFs.
There is, however, no fundamental reason why helicity PDFs would have smaller systematic effects than unpolarized ones and these smaller tensions are, again, likely the consequence of a possibly more favorable interplay of the several unaccounted for systematics.
It is still fair to assess that LQCD can have an impact on global extractions of helicity PDFs, given the smaller abundance of empirical data for polarized nucleons.
This becomes even more obvious for the transversity case (see e.g.~Refs.\ \cite{Alexandrou:2018eet,Egerer:2021dgg}), where LQCD can provide significant constraints that can complement experimental data planned to be obtained in the near future.
As argued above, the total uncertainties need to be robustly quantified to achieve this aim (see e.g.~Refs.~\cite{Alexandrou:2020qtt,Karpie:2021pap} for studies of discretization effects).
The unpolarized case can serve as a benchmark for this, given the unlikely prospect of lattice data to contribute to the global extraction of these under the abundance of empirical data taken over tens of years at various experimental facilities.

\section{New exploratory directions}
\label{sec:exploratory}
In this section, we take a short look at other recent lattice explorations of partonic distributions.
For lack of space, only subjectively chosen highlights are mentioned and for a more complete overview of the activities of the last 2 years, we refer to Refs.~\cite{Constantinou:2020pek,Cichy:2021lih}.
\begin{itemize}
\item \textbf{Flavor-singlet PDFs}.
The previous section concentrated on the case of isovector PDFs.
Getting flavor-decomposed ones is significantly more challenging, since quark-disconnected diagrams need to be computed.
These are more noisy than the quark-connected ones and thus, require more expensive computations and special techniques.
The first flavor-singlet PDFs (unpolarized, helicity and transversity) for the light quarks were obtained in Refs.~\cite{Alexandrou:2020uyt,Alexandrou:2021oih} in the quasi-PDF approach.
The disconnected diagrams were addressed via techniques optimized earlier for local operators, with explicit low modes treatment and stochastic high modes evaluation using hierarchical probing and the one-end trick.
The results were obtained with TM quarks at a non-physical pion mass ($m_\pi=260\,{\rm MeV}$) and are subject to follow-up studies of several sources of systematics.
With respect to isovector ones, an additional one is the mixing with gluon PDFs, neglected in this work.
However, this mixing can be expected to be a small effect for light quarks (but not for the heavier ones, which appear in the nucleon predominantly from gluon splitting).
\item \textbf{Gluon PDFs}.
Gluonic quantities evince significant noise in LQCD calculations and hence, gluon PDFs are difficult to determine.
The recent exploratory pseudo-distribution study by the HadStruc collaboration \cite{HadStruc:2021wmh} (clover fermions, $m_\pi=358\,{\rm MeV}$) utilized several special techniques to get a robust signal.
Particularly important was the momentum-smeared distillation technique \cite{Egerer:2020hnc}, earlier tested for quark PDFs in Ref.~\cite{Egerer:2021ymv}. 
Combining it with the summed GEVP method of extracting the matrix elements, it effectively allows to extract the data at smaller source-sink separations with demonstrable control over excited states. 
Another important technique, first introduced in Ref.~\cite{Karpie:2021pap}, was reconstruction of the $x$-distribution with Jacobi polynomials.  
At this stage, mixing with singlet quark PDFs was neglected.
The final gluon PDF was found to agree well with phenomenological determinations in the whole $x$-range.
Obviously, this cannot prevent further studies of systematics, leading to robustly quantified total uncertainties.
\item \textbf{Twist-3 PDFs}.
Apart from leading-twist (twist-2) PDFs, full characterization of nucleon's longitudinal structure requires knowledge also of higher-twist distributions.
Recently, twist-3 PDFs (chiral-even $g_T(x)$ and chiral-odd $h_L(x)$, $e(x)$) were addressed in a series of papers \cite{Bhattacharya:2020xlt,Bhattacharya:2020cen,Bhattacharya:2020jfj,Bhattacharya:2021moj}.
Two of these papers \cite{Bhattacharya:2020xlt,Bhattacharya:2020jfj} derived the 1-loop matching for the three twist-3 distributions, with special attention paid to the role of zero-mode contributions for the chiral-odd cases.
At this stage of exploratory studies, mixing with quark-gluon-quark operators was neglected (see Refs.~\cite{Braun:2021aon,Braun:2021gsk} for a discussion of taking this mixing into account).
The first lattice calculations (TM fermions, $m_\pi=260\,{\rm MeV}$) appeared in the other two articles, for $g_T$ \cite{Bhattacharya:2020cen} and $h_L$ \cite{Bhattacharya:2021moj}.
The twist-3 functions were compared to their twist-2 counterparts, the helicity and transversity PDFs.
It was observed that the twist-2 and twist-3 distributions are similar in magnitude, with a noticeably steeper descent of the latter at small $x$.
In addition, the authors performed tests of Wandzura-Wilczek (WW) approximations \cite{Wandzura:1977qf}, wherein the twist-3 functions are fully determined by their twist-2 counterparts.
Both for $g_T$ and $h_L$, the WW relations hold within statistical errors for $x\lesssim0.5$, with the magnitude of errors such that violations up to around 40\% are possible.
This still does not take into account several systematic effects, making these conclusions very preliminary.
\item \textbf{Generalized parton distributions (GPDs)}.
These off-forward generalizations of PDFs are one of the crucial distributions for probing nucleon's 3D structure.
Their first lattice determination appeared recently from ETMC -- for unpolarized ($H,\,E$) and helicity ($\tilde{H},\,\tilde{E}$) GPDs \cite{Alexandrou:2020uyt} as well as transversity ($H_T,\,E_T,\,\tilde{H}_T,\,\tilde{E}_T$) GPDs, using again TM fermions at $m_\pi=260\,{\rm MeV}$.
Results were obtained in the standard Breit frame at zero and non-zero skewness $\xi$ (when momentum transfer has a longitudinal component).
Two/four different projectors were applied to disentangle the GPDs for the chiral-even/odd case.
It was found that the GPDs are suppressed with respect to PDFs, as expected, particularly at small $x$.
For the non-zero skewness case, qualitatively different regimes were found, the DGLAP ($|x|>\xi$) and ERBL ($|x|<\xi$) regions, with GPDs further suppressed in the latter.
First preliminary results were also reported for a twist-3 GPD ($\tilde{G}_2(x)$; its combination with $\tilde{H}(x)$ being the off-forward generalization of $g_T(x)$) in the same setup \cite{Bhattacharya:2021rua}.
The GPDs calculation is significantly more complex than for PDFs and thus, the amount of work needed for their full mapping in terms of the ($x,\,t,\,\xi$) variables and the full quantification of systematics is very large. 
\item \textbf{Transverse-momentum-dependent PDFs (TMDs)}.
While GPDs incorporate information about the transverse position of partons, knowledge about their transverse motion is encoded in another class of distributions, the TMD PDFs.
In this case, the new theoretical aspect is the need to handle a new kind of divergences, so-called rapidity divergences, originating from gluon radiation and unregulated by dimensional regularization. 
A rapidity regulator can be incorporated into a (non-perturbative) soft function, which has a rapidity-independent (intrinsic) part and a part responsible for its evolution in rapidity, defining the Collins-Soper (CS) kernel.
The soft function can be obtained via LaMET, as shown in Refs.~\cite{Ji:2019sxk,Ji:2019ewn}.
On the lattice, one can calculate a pseudoscalar meson form factor, factorizable into the intrinsic soft function and a quasi-TMD wave function.
The first explorations of this method appeared recently from LPC (clover fermions, $m_\pi=333\,{\rm MeV}$ with heavier valence quarks) \cite{Zhang:2020dbb} and Beijing+ETMC (TM quarks, $m_\pi=350\,{\rm MeV}$, also with heavier valence pion mass) \cite{Li:2021wvl}, both with tree-level matching.
Feasibility of the approach was, thus, established and some first systematics were addressed (e.g.~higher-twist contamination, momentum and pion mass dependence).
Also the CS kernel calculation was recently undertaken from a ratio of quasi-TMDs at different rapidities \cite{Ebert:2018gzl} or of first Mellin moments of TMDs \cite{Schlemmer:2021aij}.
Several lattice extractions in different setups \cite{Shanahan:2020zxr,Zhang:2020dbb,Li:2021wvl,Schlemmer:2021aij,Shanahan:2021tst} lead to a qualitatively consistent picture, with the most recent one \cite{Shanahan:2021tst} concluding, unsurprisingly, an important role of power corrections and NLO perturbative matching.
Incorporating these is an important next step for the other groups, together with analyses aiming at checks of various systematic effects to establish quantitatively reliable conclusions.
\item \textbf{Other hadrons}.
The above described work concentrated on partonic distributions of the nucleon.
Obviously, there is phenomenologically relevant interest also in other hadrons.
Recent work for the pion included advanced studies aimed at estimating the systematics of such computations, using e.g.~several lattice spacings to check for discretization effects, effects of using chiral fermions, comparing renormalization procedures (including the new hybrid scheme \cite{Ji:2020brr}), assessing the role of 2-loop corrections to matching \cite{Gao:2020ito,Gao:LAT21,Zhao:LAT21}.
An attempt to reconstruct the $x$-dependence from 3 accessible low moments was made in Ref.~\cite{Alexandrou:2021mmi}.
It was found that adding the 4th moment from phenomenology has a negligible effect and also that the JAM18 pion PDF \cite{Barry:2018ort} reconstructed with 3 moments is consistent with the full one with increased uncertainties.
Hence, it is clear that higher moments become important only at a higher level of precision.
It is worth to mention that studies of pion distributions have been pursued also with other approaches: auxiliary light quark \cite{Bali:2018spj}, current-current correlators \cite{Sufian:2019bol}, pseudo-PDFs \cite{Joo:2019bzr} and auxiliary heavy quark (heavy OPE) \cite{Detmold:2021qln}, the latter recently advanced also theoretically \cite{Detmold:2021uru}.
Other than for the pion, distribution amplitudes for $K^*$ and $\phi$ mesons were recently calculated, phenomenologically important for $B/B_s$ decays wherein tensions with the Standard Model were reported by LHCb.
On the baryonic side, the PDF of the $\Delta^+$ was calculated \cite{Chai:2020nxw}, with its importance consisting in the possibility of shedding light on the sea quark asymmetry in the nucleon.
\end{itemize}

\section{Summary and prospects}
\label{sec:summary}
\vspace*{-5mm}
\begin{figure}[h!]
\begin{center} 
\hspace*{-3.5mm}
\includegraphics[scale=.46]{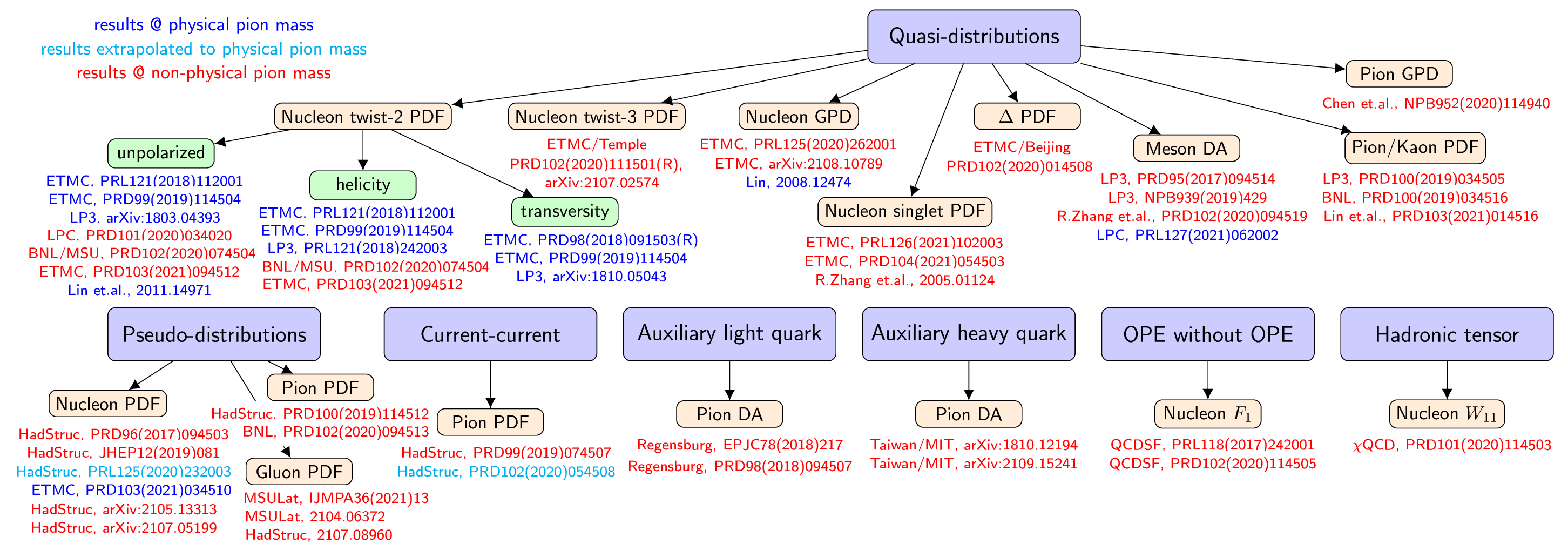}
\end{center}
\vspace*{-0.5cm}
\caption{Dynamical progress of lattice PDFs/GPDs, with papers grouped according to the method and topic.}
\label{fig:summary}
\end{figure}
The seminal 2013 paper of Ji \cite{Ji:2013dva} sparked intense interest in determining partonic distributions on the lattice, proposing a powerful framework how to access the relevant light-cone correlations with lattice-calculable Euclidean ones and also leading to the revival or conception of alternative methods, all based at some stage on factorization to get to the physical distributions.
This dynamical progress is illustrated in Fig.~\ref{fig:summary}, where papers only of the last around 3 years are referenced.

This proceeding has been written with a broader community in mind and below, we summarize the main messages that it tried to convey.
\begin{itemize}
\item Extraction of partonic distributions is difficult on the lattice, requiring multiple non-trivial ingredients.
It is crucial that control over all these ingredients is possible, but it can be achieved only with dedicated and careful studies of several sources of systematic effects.
\item Thus, the road from exploratory to precision studies is long. Most present-day calculations for partonic distributions are at the stage of exploratory or somewhat more advanced studies, but conclusions remain so far qualitative.
\item Unpolarized twist-2 PDFs, well-known from global fits to abundant sets of experimental data, are the natural benchmark for the lattice -- their reproduction within completely and properly quantified uncertainties will be one of the milestones for the lattice program.
At the present stage, there are indications that the size of systematic uncertainties is similar to their magnitude found in other lattice hadron structure studies.
Nevertheless, proper quantification of the systematics will require dedicated studies, both on the lattice (additional simulations e.g.~at several lattice spacings and volumes) and theoretical (e.g.~perturbative corrections at higher order or explicit methods to calculate power corrections).
\item For other distributions, lattice can be foreseen to provide constraints that can be incorporated into extractions from experimental data.
For several cases where experimental data are basically non-existent, lattice can provide the only available information or one that can be on a similar footing as data that will come from ongoing or planned experiments focused on nucleon's 3D structure (e.g.~from the 12 GeV program at JLab or the EIC in a more distant future).
The prerequisite for this is, again, the complete quantification of lattice uncertainties.
Under this condition, one can aim at simultaneous fits of different lattice and experimental data, treating the former as ``lattice cross sections'' in analogy with experimental ones \cite{Ma:2014jla,Ma:2017pxb}.
\end{itemize}

Overall, this points to a projected very important role for the lattice in understanding hadronic structure.
It has been shown that lattice calculations are feasible for a wide range of distributions, with varying level of difficulty, naturally increasing for more complicated distributions due to additional ingredients that need to be taken into account.
The remaining task is to make the calculations quantitatively reliable, something that is also feasible with dedicated computational and theoretical efforts.
On the lattice side, the task will be augmented by increasing computing power, necessary to handle large boosts for hadrons simulated on large and fine lattices.
As the history of lattice computations shows, optimizations in handling the simulations can likely turn out to help as well.
This allows us to conclude very optimistic prospects for this rich lattice program and eventually, predict a truly complementary role of LQCD in the quest to understand the details of the structure of the nucleon and other hadrons.\vspace*{2mm}

\begin{footnotesize}
\noindent \textbf{Acknowledgments}.
I thank the Organizers of the Virtual Tribute to Quark Confinement and the Hadron Spectrum 2021 for the invitation to give this review talk and for the very enjoyable conference.
I also thank all my Collaborators in the field of $x$-dependent partonic distributions.
Financial support from the National Science Centre (Poland) is acknowledged, grant SONATA BIS no.\ 2016/22/E/ST2/00013.
\end{footnotesize}

\bibliography{references}

\begin{thebibliography}{78}

\bibitem{Constantinou:2014tga}
M.~Constantinou, PoS \textbf{LATTICE2014}, 001 (2015), \texttt{1411.0078}

\bibitem{Liu:1993cv}
K.F. Liu, S.J. Dong, Phys. Rev. Lett. \textbf{72}, 1790 (1994),
  \texttt{hep-ph/9306299}

\bibitem{Detmold:2005gg}
W.~Detmold, C.J.D. Lin, Phys. Rev. \textbf{D73}, 014501 (2006),
  \texttt{hep-lat/0507007}

\bibitem{Braun:2007wv}
V.~Braun, D.~Mueller, Eur. Phys. J. \textbf{C55}, 349 (2008),
  \texttt{0709.1348}

\bibitem{Ji:2013dva}
X.~Ji, Phys. Rev. Lett. \textbf{110}, 262002 (2013), \texttt{1305.1539}

\bibitem{Ji:2014gla}
X.~Ji, Sci. China Phys. Mech. Astron. \textbf{57}, 1407 (2014),
  \texttt{1404.6680}

\bibitem{Lin:2014zya}
H.W. Lin, J.W. Chen, S.D. Cohen, X.~Ji, Phys. Rev. \textbf{D91}, 054510 (2015),
  \texttt{1402.1462}

\bibitem{Alexandrou:2015rja}
C.~Alexandrou, K.~Cichy, V.~Drach, E.~Garcia-Ramos, K.~Hadjiyiannakou,
  K.~Jansen, F.~Steffens, C.~Wiese, Phys. Rev. \textbf{D92}, 014502 (2015),
  \texttt{1504.07455}

\bibitem{Chen:2016utp}
J.W. Chen, S.D. Cohen, X.~Ji, H.W. Lin, J.H. Zhang, Nucl. Phys. \textbf{B911},
  246 (2016), \texttt{1603.06664}

\bibitem{Alexandrou:2016jqi}
C.~Alexandrou, K.~Cichy, M.~Constantinou, K.~Hadjiyiannakou, K.~Jansen,
  F.~Steffens, C.~Wiese, Phys. Rev. \textbf{D96}, 014513 (2017),
  \texttt{1610.03689}

\bibitem{Chambers:2017dov}
A.J. Chambers et~al., Phys. Rev. Lett. \textbf{118}, 242001 (2017),
  \texttt{1703.01153}

\bibitem{Radyushkin:2017cyf}
A.V. Radyushkin, Phys. Rev. \textbf{D96}, 034025 (2017), \texttt{1705.01488}

\bibitem{Radyushkin:2019mye}
A.~Radyushkin, Int. J. Mod. Phys. A \textbf{35}, 2030002 (2020),
  \texttt{1912.04244}

\bibitem{Ma:2014jla}
Y.Q. Ma, J.W. Qiu, Phys. Rev. \textbf{D98}, 074021 (2018), \texttt{1404.6860}

\bibitem{Ma:2017pxb}
Y.Q. Ma, J.W. Qiu, Phys. Rev. Lett. \textbf{120}, 022003 (2018),
  \texttt{1709.03018}

\bibitem{Cichy:2018mum}
K.~Cichy, M.~Constantinou, Adv. High Energy Phys. \textbf{2019}, 3036904
  (2019), \texttt{1811.07248}

\bibitem{Ji:2020ect}
X.~Ji, Y.S. Liu, Y.~Liu, J.H. Zhang, Y.~Zhao, Rev. Mod. Phys. \textbf{93},
  035005 (2021), \texttt{2004.03543}

\bibitem{Constantinou:2020pek}
M.~Constantinou, Eur. Phys. J. A \textbf{57}, 77 (2021), \texttt{2010.02445}

\bibitem{Cichy:2021lih}
K.~Cichy, \emph{{Progress in $x$-dependent partonic distributions from lattice
  QCD}}, in \emph{{38th International Symposium on Lattice Field Theory}}
  (2021), \texttt{2110.07440}

\bibitem{Green:2018vxw}
J.~Green, PoS \textbf{LATTICE2018}, 016 (2018), \texttt{1812.10574}

\bibitem{Bali:2016lva}
G.S. Bali, B.~Lang, B.U. Musch, A.~Schäfer, Phys. Rev. \textbf{D93}, 094515
  (2016), \texttt{1602.05525}

\bibitem{Constantinou:2017sej}
M.~Constantinou, H.~Panagopoulos, Phys. Rev. \textbf{D96}, 054506 (2017),
  \texttt{1705.11193}

\bibitem{Alexandrou:2017huk}
C.~Alexandrou, K.~Cichy, M.~Constantinou, K.~Hadjiyiannakou, K.~Jansen,
  H.~Panagopoulos, F.~Steffens, Nucl. Phys. \textbf{B923}, 394 (2017),
  \texttt{1706.00265}

\bibitem{Stewart:2017tvs}
I.W. Stewart, Y.~Zhao, Phys. Rev. \textbf{D97}, 054512 (2018),
  \texttt{1709.04933}

\bibitem{Izubuchi:2018srq}
T.~Izubuchi, X.~Ji, L.~Jin, I.W. Stewart, Y.~Zhao, Phys. Rev. \textbf{D98},
  056004 (2018), \texttt{1801.03917}

\bibitem{Orginos:2017kos}
K.~Orginos, A.~Radyushkin, J.~Karpie, S.~Zafeiropoulos, Phys. Rev.
  \textbf{D96}, 094503 (2017), \texttt{1706.05373}

\bibitem{Ji:2020brr}
X.~Ji, Y.~Liu, A.~Sch\"afer, W.~Wang, Y.B. Yang, J.H. Zhang, Y.~Zhao, Nucl.
  Phys. B \textbf{964}, 115311 (2021), \texttt{2008.03886}

\bibitem{Zhang:2020rsx}
K.~Zhang, Y.Y. Li, Y.K. Huo, A.~Sch\"afer, P.~Sun, Y.B. Yang
  (\ensuremath{\chi}QCD), Phys. Rev. D \textbf{104}, 074501 (2021),
  \texttt{2012.05448}

\bibitem{LatticePartonCollaborationLPC:2021xdx}
Y.K. Huo et~al. (Lattice Parton Collaboration (LPC)), Nucl. Phys. B
  \textbf{969}, 115443 (2021), \texttt{2103.02965}

\bibitem{Karpie:2019eiq}
J.~Karpie, K.~Orginos, A.~Rothkopf, S.~Zafeiropoulos, JHEP \textbf{04}, 057
  (2019), \texttt{1901.05408}

\bibitem{Braun:2020ymy}
V.~Braun, K.~Chetyrkin, B.~Kniehl, JHEP \textbf{07}, 161 (2020),
  \texttt{2004.01043}

\bibitem{Chen:2020arf}
L.B. Chen, W.~Wang, R.~Zhu, Phys. Rev. D \textbf{102}, 011503 (2020),
  \texttt{2005.13757}

\bibitem{Chen:2020iqi}
L.B. Chen, W.~Wang, R.~Zhu, JHEP \textbf{10}, 079 (2020), \texttt{2006.10917}

\bibitem{Chen:2020ody}
L.B. Chen, W.~Wang, R.~Zhu, Phys. Rev. Lett. \textbf{126}, 072002 (2021),
  \texttt{2006.14825}

\bibitem{Li:2020xml}
Z.Y. Li, Y.Q. Ma, J.W. Qiu, Phys. Rev. Lett. \textbf{126}, 072001 (2021),
  \texttt{2006.12370}

\bibitem{Bhat:2020ktg}
M.~Bhat, K.~Cichy, M.~Constantinou, A.~Scapellato, Phys. Rev. D \textbf{103},
  034510 (2021), \texttt{2005.02102}

\bibitem{Alexandrou:2018pbm}
C.~Alexandrou, K.~Cichy, M.~Constantinou, K.~Jansen, A.~Scapellato,
  F.~Steffens, Phys. Rev. Lett. \textbf{121}, 112001 (2018),
  \texttt{1803.02685}

\bibitem{Alexandrou:2019lfo}
C.~Alexandrou, K.~Cichy, M.~Constantinou, K.~Hadjiyiannakou, K.~Jansen,
  A.~Scapellato, F.~Steffens, Phys. Rev. \textbf{D99}, 114504 (2019),
  \texttt{1902.00587}

\bibitem{Joo:2020spy}
B.~Jo\'o, J.~Karpie, K.~Orginos, A.V. Radyushkin, D.G. Richards,
  S.~Zafeiropoulos, Phys. Rev. Lett. \textbf{125}, 232003 (2020),
  \texttt{2004.01687}

\bibitem{Cichy:2019ebf}
K.~Cichy, L.~Del~Debbio, T.~Giani, JHEP \textbf{10}, 137 (2019),
  \texttt{1907.06037}

\bibitem{DelDebbio:2020rgv}
L.~Del~Debbio, T.~Giani, J.~Karpie, K.~Orginos, A.~Radyushkin,
  S.~Zafeiropoulos, JHEP \textbf{02}, 138 (2021), \texttt{2010.03996}

\bibitem{Bringewatt:2020ixn}
J.~Bringewatt, N.~Sato, W.~Melnitchouk, J.W. Qiu, F.~Steffens, M.~Constantinou,
  Phys. Rev. D \textbf{103}, 016003 (2021), \texttt{2010.00548}

\bibitem{Alexandrou:2018eet}
C.~Alexandrou, K.~Cichy, M.~Constantinou, K.~Jansen, A.~Scapellato,
  F.~Steffens, Phys. Rev. \textbf{D98}, 091503 (2018), \texttt{1807.00232}

\bibitem{Egerer:2021dgg}
C.~Egerer et~al. (2021), \texttt{2111.01808}

\bibitem{Alexandrou:2020qtt}
C.~Alexandrou, K.~Cichy, M.~Constantinou, J.R. Green, K.~Hadjiyiannakou,
  K.~Jansen, F.~Manigrasso, A.~Scapellato, F.~Steffens, Phys. Rev. D
  \textbf{103}, 094512 (2021), \texttt{2011.00964}

\bibitem{Karpie:2021pap}
J.~Karpie, K.~Orginos, A.~Radyushkin, S.~Zafeiropoulos (2021),
  \texttt{2105.13313}

\bibitem{Alexandrou:2020uyt}
C.~Alexandrou, M.~Constantinou, K.~Hadjiyiannakou, K.~Jansen, F.~Manigrasso,
  Phys. Rev. Lett. \textbf{126}, 102003 (2021), \texttt{2009.13061}

\bibitem{Alexandrou:2021oih}
C.~Alexandrou, M.~Constantinou, K.~Hadjiyiannakou, K.~Jansen, F.~Manigrasso,
  Phys. Rev. D \textbf{104}, 054503 (2021), \texttt{2106.16065}

\bibitem{HadStruc:2021wmh}
T.~Khan et~al. (HadStruc) (2021), \texttt{2107.08960}

\bibitem{Egerer:2020hnc}
C.~Egerer, R.G. Edwards, K.~Orginos, D.G. Richards, Phys. Rev. D \textbf{103},
  034502 (2021), \texttt{2009.10691}

\bibitem{Egerer:2021ymv}
C.~Egerer, R.G. Edwards, C.~Kallidonis, K.~Orginos, A.V. Radyushkin, D.G.
  Richards, E.~Romero, S.~Zafeiropoulos (2021), \texttt{2107.05199}

\bibitem{Bhattacharya:2020xlt}
S.~Bhattacharya, K.~Cichy, M.~Constantinou, A.~Metz, A.~Scapellato,
  F.~Steffens, Phys. Rev. D \textbf{102}, 034005 (2020), \texttt{2005.10939}

\bibitem{Bhattacharya:2020cen}
S.~Bhattacharya, K.~Cichy, M.~Constantinou, A.~Metz, A.~Scapellato,
  F.~Steffens, Phys. Rev. D \textbf{102}, 111501 (2020), \texttt{2004.04130}

\bibitem{Bhattacharya:2020jfj}
S.~Bhattacharya, K.~Cichy, M.~Constantinou, A.~Metz, A.~Scapellato,
  F.~Steffens, Phys. Rev. D \textbf{102}, 114025 (2020), \texttt{2006.12347}

\bibitem{Bhattacharya:2021moj}
S.~Bhattacharya, K.~Cichy, M.~Constantinou, A.~Metz, A.~Scapellato, F.~Steffens
  (2021), \texttt{2107.02574}

\bibitem{Braun:2021aon}
V.M. Braun, Y.~Ji, A.~Vladimirov, JHEP \textbf{05}, 086 (2021),
  \texttt{2103.12105}

\bibitem{Braun:2021gsk}
V.M. Braun, Y.~Ji, A.~Vladimirov (2021), \texttt{2108.03065}

\bibitem{Wandzura:1977qf}
S.~Wandzura, F.~Wilczek, Phys. Lett. \textbf{72B}, 195 (1977)

\bibitem{Bhattacharya:2021rua}
S.~Bhattacharya, K.~Cichy, M.~Constantinou, A.~Metz, A.~Scapellato,
  F.~Steffens, \emph{{Twist-3 partonic distributions from lattice QCD}}, in
  \emph{{28th International Workshop on Deep Inelastic Scattering and Related
  Subjects}} (2021), \texttt{2107.12818}

\bibitem{Ji:2019sxk}
X.~Ji, Y.~Liu, Y.S. Liu, Nucl. Phys. B \textbf{955}, 115054 (2020),
  \texttt{1910.11415}

\bibitem{Ji:2019ewn}
X.~Ji, Y.~Liu, Y.S. Liu, Phys. Lett. B \textbf{811}, 135946 (2020),
  \texttt{1911.03840}

\bibitem{Zhang:2020dbb}
Q.A. Zhang et~al. (Lattice Parton), Phys. Rev. Lett. \textbf{125}, 192001
  (2020), \texttt{2005.14572}

\bibitem{Li:2021wvl}
Y.~Li et~al. (2021), \texttt{2106.13027}

\bibitem{Ebert:2018gzl}
M.A. Ebert, I.W. Stewart, Y.~Zhao, Phys. Rev. \textbf{D99}, 034505 (2019),
  \texttt{1811.00026}

\bibitem{Schlemmer:2021aij}
M.~Schlemmer, A.~Vladimirov, C.~Zimmermann, M.~Engelhardt, A.~Sch\"afer, JHEP
  \textbf{08}, 004 (2021), \texttt{2103.16991}

\bibitem{Shanahan:2020zxr}
P.~Shanahan, M.~Wagman, Y.~Zhao, Phys. Rev. D \textbf{102}, 014511 (2020),
  \texttt{2003.06063}

\bibitem{Shanahan:2021tst}
P.~Shanahan, M.~Wagman, Y.~Zhao (2021), \texttt{2107.11930}

\bibitem{Gao:2020ito}
X.~Gao, L.~Jin, C.~Kallidonis, N.~Karthik, S.~Mukherjee, P.~Petreczky,
  C.~Shugert, S.~Syritsyn, Y.~Zhao, Phys. Rev. D \textbf{102}, 094513 (2020),
  \texttt{2007.06590}

\bibitem{Gao:LAT21}
X.~Gao et~al., PoS \textbf{LATTICE2021}, 598 (2021)

\bibitem{Zhao:LAT21}
Y.~Zhao et~al., PoS \textbf{LATTICE2021}, 614 (2021)

\bibitem{Alexandrou:2021mmi}
C.~Alexandrou, S.~Bacchio, I.~Clo\"et, M.~Constantinou, K.~Hadjiyiannakou,
  G.~Koutsou, C.~Lauer (ETM), Phys. Rev. D \textbf{104}, 054504 (2021),
  \texttt{2104.02247}

\bibitem{Barry:2018ort}
P.~Barry, N.~Sato, W.~Melnitchouk, C.R. Ji, Phys. Rev. Lett. \textbf{121},
  152001 (2018), \texttt{1804.01965}

\bibitem{Bali:2018spj}
G.S. Bali et~al., Phys. Rev. \textbf{D98}, 094507 (2018), \texttt{1807.06671}

\bibitem{Sufian:2019bol}
R.S. Sufian, J.~Karpie, C.~Egerer, K.~Orginos, J.W. Qiu, D.G. Richards, Phys.
  Rev. \textbf{D99}, 074507 (2019), \texttt{1901.03921}

\bibitem{Joo:2019bzr}
B.~Joó, J.~Karpie, K.~Orginos, A.V. Radyushkin, D.G. Richards, R.S. Sufian,
  S.~Zafeiropoulos, Phys. Rev. \textbf{D100}, 114512 (2019),
  \texttt{1909.08517}

\bibitem{Detmold:2021qln}
W.~Detmold, A.~Grebe, I.~Kanamori, C.J.D. Lin, S.~Mondal, R.~Perry, Y.~Zhao
  (2021), \texttt{2109.15241}

\bibitem{Detmold:2021uru}
W.~Detmold, A.V. Grebe, I.~Kanamori, C.J.D. Lin, R.J. Perry, Y.~Zhao (HOPE),
  Phys. Rev. D \textbf{104}, 074511 (2021), \texttt{2103.09529}

\bibitem{Chai:2020nxw}
Y.~Chai et~al., Phys. Rev. D \textbf{102}, 014508 (2020), \texttt{2002.12044}

\end{thebibliography}

\end{document}